\documentclass[twocolumn,preprintnumbers,aps,nofootinbib,amssymb,amsmath,showpacs,showkeys]{revtex4}
\usepackage{amssymb}
\usepackage{dsfont}
\usepackage{amsmath}

\newcommand{\bastar}{\begin{eqnarray*}}
\newcommand{\eastar}{\end{eqnarray*}}

\newcommand{\be}{\begin{equation}}
\newcommand{\ee}{\end{equation}}
\newcommand{\ba}{\begin{array}}
\newcommand{\ea}{\end{array}}
\newcommand{\bea}{\begin{eqnarray}}
\newcommand{\eea}{\end{eqnarray}}
\newcommand{\pro}{\partial}

\newcommand{\difrac}{\displaystyle\frac}
\newcommand{\nn}{\nonumber}

\newcommand{\starq}{{\stackrel{*}{q}}}

\newcommand{\TS}{\stackrel{\top}{S}}
\newcommand{\TTS}{\stackrel{\top\!\!\top}{S}}
\newcommand{\TR}{\stackrel{\top}{R}}
\newcommand{\TTR}{\stackrel{\top\!\!\top}{R}}
\newcommand{\oD}{{\cal D}}

\newcommand{\und}{\underline}

\makeatletter
\def\eqnarray{ \stepcounter{equation} \let\@currentlabel=\theequation
\global\@eqnswtrue
\global\@eqcnt\z@
\tabskip\@centering
\let\\=\@eqncr
$$\halign to \displaywidth\bgroup\@eqnsel\hskip\@centering
$\displaystyle\tabskip\z@{##}$&\global\@eqcnt\@ne
\hfil$\displaystyle{{}##{}}$\hfil
&\global\@eqcnt\tw@$\displaystyle\tabskip\z@{##}$\hfil
\tabskip\@centering&\llap{##}\tabskip\z@\cr}
\makeatother

\makeatletter
\def\@arrayacol{\edef\@preamble{\@preamble \hskip .5\arraycolsep}}
\def\array{\let\@acol\@arrayacol \let\@classz\@arrayclassz
\let\@classiv\@arrayclassiv \let\\\@arraycr\def\@halignto{}\@tabarray}
\makeatother

\renewcommand{\arraystretch}{1.6}

\begin{document}

\title{
Lorentz gauge theory as a model of emergent gravity}
\bigskip
\author{D. G. Pak}
\email{dmipak@gmail.com}
\affiliation{
Institute of Modern Physics of CAS, Lanzhou 730000, China}
\affiliation{Lab. of Few Nucleon Systems, Institute for Nuclear Physics, Ulughbek, 100214, Uzbekistan}
\author{Youngman Kim}
\email{ykim@apctp.org}
\affiliation{
Asia Pacific Center for Theoretical Physics, Pohang, Gyeongbuk 790-784, Korea}
\affiliation{Department of Physics, Pohang University of Science and Technology,
Pohang, Gyeongbuk 790-784, Korea}
\author{Takuya Tsukioka}
\email{tsukioka@apctp.org}
\affiliation{
Asia Pacific Center for Theoretical Physics, Pohang, Gyeongbuk 790-784, Korea}


\begin{abstract}
We consider a class of Lorentz gauge gravity theories
within Riemann-Cartan geometry which admits a topological
phase in the gravitational sector. The dynamic content of such theories
is determined only by the contortion part of the Lorentz gauge connection.
We demonstrate that there is a unique Lagrangian that
admits propagating spin one mode in correspondence
with gauge theories of other fundamental interactions.
Remarkably, despite the $R^2$ type of the Lagrangian
and non-compact structure of the Lorentz gauge group, the model
possesses rather a positive-definite Hamiltonian.
This has been proved in the lowest order of perturbation theory.
This implies further consistent quantization and leads
to renormalizable quantum theory. It is assumed that the proposed
model describes possible mechanism of emergent Einstein gravity
at very early stages of the Universe due to quantum dynamics of
contortion.
\end{abstract}
\vspace{0.3cm}
\maketitle

\section{Introduction}

The idea that Lorentz gauge approach can lead to a consistent
quantum theory of gravity has been developed for last
fifty years since the seminal paper by Utiyama \cite{uti}.
The exhausting list of references can be found in
reviews on this topic (see, for instance,
\cite{hehl,ivan1}). Among early works devoted to
Lorentz gauge theory with Yang-Mills type Lagrangian
one should mention the
papers \cite{carmeli1,carmeli5,lord,martell,anttomb,john}
where main features of classical and quantum theory were studied.
Extension of the Lorentz gauge approach to the case of general
Lorentz connection including contortion was widely explored
as well \cite{hehl,ivan1,odints,shapiro}.
The most general Lagrangian quadratic in Riemann-Cartan curvature
and with Einstein-Hilbert term was considered in \cite{hayashiI}.
Recently a Lorentz gauge gravity model with contortion part in the
Lorentz gauge connection has been proposed
\cite{IJMPA2010} which admits a topological phase for gravitation.
We assume that such a topological phase can be possibly realized
at very early stages of our Universe close to or before the Bing Bang.
The standard gravity supposed
to be an effective theory which is induced during phase transition
due to quantum dynamics of contortion. The idea that Einstein gravity is
an effective theory and can be deduced from some more fundamental
theory is not new, it was sounded by Zel'dovich and Sakharov
in 70s \cite{zeld, sakhar}.
Possible mechanisms of inducing the Einstein theory
via quantum corrections were proposed in past by many physicists
in various approaches: conformal invariance breaking schemes
\cite{adler1,adler2}, non-linear realizations of the Lorentz group
\cite{tseytlin,leclerc}, models with spontaneous symmetry
breaking \cite{ogiev,isham1,zee,kirs,pirog},
superstring models, loop quantum gravity \cite{rovelli,smolin}
and others \cite{macdow,pereira}.
In order to capture the nature of gravity,
thermodynamic approaches have been also developed~\cite{jac,pad,kps}.
Recently, it was conjectured that the gravity could be regarded as
the entropic force through the holographic principle~\cite{ver}.
In most of these approaches the Einstein-Hilbert term is induced by
quantum corrections due to interaction with matter field.

Our approach is based on the gauge principle
which was successfully realized in formulating the theories of
electro-weak and strong interactions.
We consider the local Lorentz symmetry as an appropriate gauge symmetry
for constructing a generalized theory of gravity in geometric framework
since it reflects the equivalence principle,
which is a corner stone of general relativity.
This introduces naturally the contortion as a part of general
Lorentz gauge connection.
Whether or not the contortion (torsion) is relevant to our real world
is discussed in detail in \cite{hehlobukh}.

We consider theories with a Lagrangian containing only Riemann-Cartan
curvature squared terms. We do not introduce terms quadratic in torsion since
we treat the contortion as a part of Lorentz gauge connection,
not as a tensor. By this way we keep the gauge structure of the considered
Lorentz gauge gravity models close to standard gauge approach.
It has been shown \cite{IJMPA2010} that there is a model
with a special $R^2$ type Lagrangian which admits a topological phase
for the gravitation whereas contortion still possesses dynamical
degrees of freedom. An interesting feature of the model is
that the number of dynamical degrees of freedom of torsion is the same
as the number of physical degrees of the metric tensor.
This gives a hint that torsion may play a role of quantum counter
part to the classical metric of Einstein gravity
which supposed to be an effective theory generated by the quantum dynamics
of torsion \cite{CQG2008}. The analysis of dynamic content of the model
in \cite{IJMPA2010} has been performed at the lowest linearized level
in contortion part and
in the presence of constant Riemann curvature space-time background.
Due to these limitations several important
issues in this model remain unclear, especially,
whether the dynamical properties of torsion are intrinsic properties or
they depend on presence of the background metric.

As it is known, theories with $R^2$ type Lagrangian
suffer from a serious problem related to non-definiteness
of the Hamiltonian due to non-compact structure of the Lorentz gauge
group. This has been the main obstacle toward consistent quantization
and defining a physical unitary $S$ matrix. One possible way to overcome
this problem is based on Euclidean gravity formalism \cite{hawking,hamber,hamber2011}.
One should notice, that presence of higher derivative terms
in the Lagrangian still may cause problems with unitarity
and ghosts in the graviton propagator in Euclidean gravity \cite{hamber, salam-str78}.

In the present paper we study dynamical properties of the
topological gravity model with torsion
in the limit of flat space-time metric.
We have found that Lorentz gauge
connection has dynamic degrees of freedom with a
Lagrangian specified by the same set of
parameters in the initial Lagrangian as in the case of
the presence of background constant Riemannian curvature space-time.
This proves that contortion possesses genuine dynamical
properties independently on the metric.
It is unexpected, we have demonstrated in the lowest order of perturbation
theory that the model has a positive definite Hamiltonian.
This allows to define stable quantum vacuum and perform consistent
quantization preserving unitarity in the theory.

In Section II we present the principal ideas
lying in the basis of the model of quantum gravity with contortion.
In Section III we study the dynamic content
of the theory by solving equations of motion
in Lagrange formalism. All equations of motion
are solved in linearized approximation by using decomposition
of the Lorentz connection around fixed classical solution corresponding to
constant torsion background. In Section IV we prove the positive definiteness
of the Hamiltonian in the linearized approximation.
The last section contains discussion of possible physical implications.

\section{Lorentz gauge theory with topological gravity}

Lorentz gauge theory on curved space-time can be described naturally
within Riemann-Cartan geometrical formalism.
Let us start first with the main outlines of Riemann-Cartan geometry.
The basic geometric objects  are
the vielbein $e_a^m$ and the general Lorentz affine connection $A_{m cd}$
which can be identified with the Lorentz gauge potential.
The infinitesimal Lorentz transformation of the
vielbein $e_a^m$ is given by
\be
\delta e_a^m= \Lambda_a{}^b e_b^m,
\ee
where $\Lambda_{ab}\ (=-\Lambda_{ba})$ is the Lorentz gauge parameter.
We use $m,n,\ldots$ to denote world indices,
and $a,b,\ldots$ for Lorentz frame indices. We assume that the vielbein is
invertible and the metric $\eta_{ab} \ (=e_a^me_{mb})$ has
Lorentz signature $\eta_{ab}={\rm diag} (-,+,+,+)$.

The covariant derivative with respect to the Lorentz group
transformation is defined in a standard manner
\be
D_a=e_a^m (\pro_m + g{\bf A}_m) ,
\ee
where ${\bf A}_m\equiv A_{m cd} \Omega^{cd}$ is affine connection
taking values in the Lorentz Lie algebra whose generator is given by $\Omega^{cd}$,
and $g$ is a new gravitational gauge coupling constant.
For brevity of notation we will use a
redefined connection which absorbs the coupling constant. The
original Lorentz gauge transformation of the connection ${\bf A}_m$ has
the form
\be
 \delta {\bf A}_m=-\pro_m {\bf \Lambda} - [{\bf A}_m, {\bf \Lambda}],
\label{eq:deltaA}
\ee
 where ${\bf \Lambda}=\Lambda_{cd}\Omega^{cd}$.
The Lorentz gauge connection $A_{m a}{}^b$ can be rewritten
as the sum
\be
A_{m ab}
= \varphi_{m ab}(e) + K_{m ab},
\label{split 1}
\ee
where $K_{m ab}$ is a contortion and $\varphi_{m ab}(e) $ is a
Levi-Civita spin connection given in terms of the vielbein
\be
\varphi_{m ab}(e)
=
-\difrac{1}{2}
\Big(e^n_b \pro_m e_{n a}-
e_a^n e_m^c \pro_b e_{n c}+\pro_a e_{mb}
-(a\leftrightarrow b)
\Big).
\ee
The torsion and curvature tensors are defined in a standard way
\be
[D_a,D_b]=T_{ab}^{\,\,c} D_c+R_{ab cd} \Omega^{cd},
\ee
where the torsion components
in the unholonomic basis can be expressed in terms of contortion, and conversely
\be
\begin{array}{rcl}
T_{ab}^{\,\,c}
&=&K_{ab}^{~c}-K_{ba}^{~c}, \\
K_{abc}
&=&\dfrac{1}{2}(T_{abc}-T_{bca}+T_{cab}).
\end{array}
\ee

The most general quadratic in Riemann-Cartan curvature Lagrangian reads
\bea
{\cal L}
&=&
c_1 R_{ab cd} R^{abcd}+c_2R_{ab cd} R^{cdab}+c_3 R_{ab} R^{ab}
\nonumber
\\
&&
+c_4 R_{ab} R^{ba}
+ c_5 R^2 + c_6 A_{abcd}^2, \label{Lgeneral}
\eea
where the last term is an additional invariant
which appears in Riemann-Cartan space-time.
The tensor $A_{abcd}$ is defined as follows \cite{hayashiI}
\be
A_{abcd}
\equiv
\dfrac{1}{6} (R_{abcd}+R_{acdb}+R_{adbc}+R_{bcad}+R_{bdca}+R_{cdab}).
\ee
In Riemannian space-time the tensor $A_{abcd}$ vanishes due to the
Jacobi cyclic identity
\be
R_{abcd}+R_{acdb}+R_{adbc}= 0 .
\ee
A careful analysis of gravity models including Einstein term in the Lagrangian
was done in \cite{hayashiI}. We do not consider Einstein term since
we treat the Einstein gravity as an effective theory which should not
be quantized and which is induced from a more general theory, in our case
from Riemann-Cartan gravity. So that, only contortion represents
quantum dynamical degree of freedom in a special Riemann-Cartan gravity model.
In general the Lagrangian (\ref{Lgeneral}) contains propagating
modes for both fields, metric and contortion. So that,
formally the metric can still be considered
as a quantum field as well as the contortion.
This is not merely satisfactory because metric and contortion
represent different geometric objects. The metric plays a role
of kinematic variable in description of the space-time geometry,
whereas the contortion, as a part of gauge connection, plays a
role of gauge potential which represents dynamic object in
gauge theories of electroweak and strong interactions.
To keep only the contortion as a quantum variable we conjecture
that a generalized Riemann-Cartan gravity may admit a phase where
the metric describes a pure topological structure of the space-time.
So that the metric does not satisfy any equations of motion and it
cannot be quantized in principle. This is our main idea.
We are looking for such a Lagrangian in Riemann-Cartan space-time
which reduces to topological Gauss-Bonnet density in the limit
of Riemannian geometry.

In Riemann-Cartan geometry
the proper generalization of the topological Gauss-Bonnet invariant
(Euler characteristic) is given by
the Bach-Lanczos density \cite{bach-lanczos,lanczos}
\be
I_{\rm BL} =R_{ab cd} R^{cdab}-4 R_{ab} R^{ba} + R^2.
\ee
 The properties of the Bach-Lanczos invariant
are described in a detail in \cite{hayashiV}.
A proper Lagrangian can be derived from the general expression
(\ref{Lgeneral}) by fitting the parameters $c_i$
as follows
\bea
{\cal L}
&=&
-\dfrac{1}{32}
\Big\{
\alpha R_{abcd}^2+(1-\alpha) R_{abcd}R^{cdab} -4 \beta R_{bd}^2
\nonumber
\\
&&
\hspace*{9mm}
-4(1-\beta) R_{bd}R^{db} +R^2+6 \gamma A_{abcd}^2
\Big\},
\eea
where the parameters $\alpha,\beta, \gamma$ remain arbitrary.
One can check that the
Lagrangian reduces to the Gauss-Bonnet density
in the limit of Riemannian space-time, i.e.,
when contortion is set to be zero.
One can rewrite the Lagrangian in a more simple form
\bea
{\cal L}
&=&
-\dfrac{1}{32}
\Big\{
(\alpha+\gamma)R_{abcd}^2
-(\alpha-\gamma)R_{abcd}R^{cdab}
\nonumber
\\
&&
\hspace*{9mm}
+ 4 \gamma R_{abcd}R^{acdb}
-4 \beta( R_{bd}^2-R_{bd}R^{db})
\nonumber
\\
&&
\hspace*{9mm}
+I_{\rm BL}
\Big\}\label{startL}.
 \eea
It has been shown that the model described by the Lagrangian (\ref{startL})
admits dynamical degrees of freedom for the contortion only for
the special values of the parameters, $\beta=0, \gamma=-3\alpha$
with overall normalization factor $\alpha$ \cite{IJMPA2010}.
The result has been obtained from
the analysis of linearized equations of motion for contortion
in the presence of constant Riemann curvature space-time background.
Therefore, the principal question arises whether contortion will
keep its properties in the flat Riemannian space-time.
In other words, whether the dynamics of torsion represents
its intrinsic properties independent of the metric. If the
contortion still possesses dynamical properties in flat space-time,
then another important question arises,
at which values of the parameters $\alpha,\beta, \gamma$
it will happen.

In the present paper we will mainly concentrate on flat metric limit
i.e.\ a pure Lorentz gauge theory
with the Lagrangian of type (\ref{startL}).
The field strength (curvature tensor) in flat space-time takes a simple form
\be
R_{mncd}= \pro_m A_{ncd}+A_{mce} A_{ned} - (m\leftrightarrow n).
\ee
 Since the background vielbein is flat there is no difference between
the world and Lorentzian indices.
Our study will be constrained by a special choice of the parameter,
$\beta=0$, which is a necessary condition of existence of propagating
vector mode in the presence of constant curvature
space-time \cite{IJMPA2010}.

The theory described by the Lagrangian (\ref{startL}) is highly non-linear
and belongs to degenerate theories \cite{gitman}. Application of canonical
formalism to such theories is quite complicated due to the presence of
constraints of higher orders. Therefore, to
study the dynamical structure of the theory we will use Lagrange formalism
and apply linearized approximation method which is effective in the
analysis of non-linear equations of motion.
We will split the Lorentz gauge connection
into classical background field $B_{acd}$ (which plays a role
of the mean field) and fluctuating part $q_{acd}$ as follows
\be
A_{a cd}= B_{a cd} + q_{a cd}. \label{split}
\ee
Under the decomposition (\ref{split})
the general field strength is split into two parts as follows
\bea
R_{abcd}&=&{\cal R}_{abcd}(B)+\tilde R_{abcd}(q), \nn \\
 {\cal R}_{abcd}(B)
&=&\pro_a B_{bcd}+B_{ace} B_{bed} - (a\leftrightarrow b),  \\
\tilde R_{abcd}(q)
&=&{\cal D}_{\underline a} q_{b \underline c \underline d}+q_{ace} q_{bed}
- (a\leftrightarrow b), \nn
\eea
where ${\cal D}_a$ is a background covariant derivative containing
the classical field $B_{acd}$,
and the underlined indices stand for indices over which the
covariantization is performed.

There are two gauge non-equivalent representations for
gauge potentials leading to the same constant field strength
in $SU(2)$ Yang-Mills theory:
Abelian type and non-Abelian type \cite{brown,leut,schwab}.
In the case of constant curvature space-time
the Abelian type of gravitational field has been used
for spin connection \cite{IJMPA2010}. The calculations
are crucially simplified using normal coordinate decomposition of the metric.
In the present case of flat space-time it is more
convenient to choose a constant background field of non-Abelian type
defined by the following Lorentz gauge potential $B_{acd}$,
\bea
B_{0cd}&=&0, \nn \\
B_{\alpha\beta\gamma}&=&\epsilon_{\alpha\beta\gamma} H,
\label{backsol}\\
B_{\alpha 0 \beta}&=&\delta_{\alpha \beta} G,  \nn
\eea
where Greek indices run through the space components
and $\epsilon_{123}=+1$.
The constant field is determined by two number parameters
$G,H$ which correspond to rank two of the Lorentz group.
The corresponding field strength reads
\bea
{\cal R}_{0\alpha cd}
&=&0, \nn \\
{\cal R}_{\alpha\beta 0\delta}
&=&-2 \epsilon_{\alpha\beta\delta} HG, \\
{\cal R}_{\alpha\beta\gamma\delta}
&=&(H^2-G^2) (\delta_{\alpha\gamma} \delta_{\beta\delta}
-\delta_{\alpha\delta}\delta_{\beta\gamma}). \nn
\eea
We will analyze the equations of motion in a detail for the case
of constant background $G=0, H \neq 0$ which is
one of the background solutions.

\section{Equations of motion in Lagrange formalism}

The classical theory with the Lagrangian (\ref{startL}) is
degenerate. This implies that the number of equations of motion
in free theory is less than number of field degrees of freedom.
So that, one has to consider non-linear equations of motion
to determine the dynamic content of all fields.
The degeneracy of the quadratic Lagrangian (\ref{startL})
manifests in appearance of additional local symmetries.
One symmetry is similar to $U(1)$ gauge symmetry
\be
\begin{array}{rcl}
\delta_{U(1)} q_{acd}
&=& \dfrac{1}{3} (\eta_{ac}\pro_d\lambda-\eta_{ad}\pro_c \lambda), \\
\delta_{U(1)} q^a{}_{ad}
&=& \pro_d \lambda,
\end{array}
\label{U1}
\ee
and it implies that only transverse degrees of freedom
of the vector field $q^a{}_{ad}$
can be propagating. Another symmetry with a constrained parameter
$\chi_{bc}$ has the following form
\be
\delta_\chi q_{acd}= \pro_c \chi_{da}-\pro_d \chi_{ca},
\label{eqchi}
\ee
where
$\chi_{bc}=\chi_{cb}, \chi^c_{~c}=0 $ and $\pro^c\chi_{cd}=0$.
These symmetries reduce essentially the number of dynamical component
fields in the contortion.

Let us consider linearized equations of motion corresponding
to the Lagrangian (\ref{startL})
\bea
\dfrac{\delta {\cal L}}{\delta q_{ncd}}
&\equiv&
(\alpha+\gamma)\oD^{\und m} \big(\oD_{m} q_{n {\und c} {\und d}}
-\oD_{n} q_{m {\und c} {\und d}}\big)
\nn
\\
&&
-(\alpha-\gamma)\oD^{\und m}\big(\oD_{ {\und c}} q_{\und d  m  n}
-\oD_{ {\und d}} q_{\und c  m  n}\big) \nn \\
&&
+\gamma \oD^{\und m} \Big(\oD_{ m} q_{\und c {\und d} n}
-\oD_{ {\und c}} q_{m  {\und d} n}
\nn
\\
&&
\hspace*{13mm}
-\oD_{ n} q_{\und c {\und d}  m}
+\oD_{ {\und c}} q_{n  {\und d} m}-(c\leftrightarrow d)\Big)
\nn
\\
&=&0,
\label{eqcov}
\eea
where covariant derivatives inside the brackets
act on the last two indices of $q_{ncd}$,
and for the second covariant derivatives,
$\oD^{\und m}$, the covariantization is performed over underlined indices.
One has twenty four equations of motion, six equations among them represent
Noether identities due to local Lorentz symmetry.
One has to impose six gauge fixing
conditions which will be chosen in consistence with equations of motion.

It is convenient to
make the following decomposition of the Lorentz gauge connection
$q_{a cd}$ into irreducible parts
$(q_{00\mu}, q_{0\mu\nu}, q_{\mu\gamma\delta}, q_{\mu 0\rho})$
where
\be
\renewcommand{\arraystretch}{1.9}
\begin{array}{rcl}
q_{\mu\gamma\delta}
&=&
\epsilon_{\gamma \delta\rho}
\Big(\TTS_{\mu\rho}+\dfrac{1}{2}(\delta_{\mu\rho}
-\dfrac{\pro_\mu\pro_\rho}{\Delta})\TS
\\
&&
\hspace*{9mm}
 +(\pro_\mu S_\rho+\pro_\rho S_\mu)
+\epsilon_{\mu\rho\sigma} A_\sigma
\Big),
\\
q_{\mu 0 \rho}
&=&
\TTR_{\mu\rho}+\dfrac{1}{2}(\delta_{\mu\rho}
-\dfrac{\pro_\mu\pro_\rho}{\Delta})\TR
\\
&&
 +(\pro_\mu R_\rho+\pro_\rho R_\mu) +\epsilon_{\mu\rho\sigma} Q_\sigma.
\end{array}
\renewcommand{\arraystretch}{1.6}
\ee
We define $\Delta=\pro_\alpha \pro_\alpha$, and
the superscript ``$\top$'' stands for traceless components
and ``$\top\!\!\top$''
denotes traceless and transverse irreducible part.
The decomposition is similar to that used for the metric tensor
in canonical formalism of Einstein gravity \cite{deser}.
Note that the fields $\TS, \TR$ and longitudinal components
$A_\alpha^{\rm l}=\frac{\pro_\alpha\pro_\beta}{\Delta}A_\beta,
Q_\alpha^{\rm l}=\frac{\pro_\alpha\pro_\beta}{\Delta}Q_\beta$
do not transform under Lorentz gauge transformations.

We will solve all equations of motion in component form.
Let us start with the equation
\bea
\dfrac{\delta {\cal L}}{\delta q_{00\delta}}
&\equiv&
\Delta q_{00\delta}-\pro_\mu \pro_\delta q_{00\mu}+
   \pro_\mu \pro_0 (q_{\delta 0 \mu}-q_{\mu 0 \delta}) \nn \\
&&
+4 H \epsilon_{\mu\delta \epsilon} \pro_\mu q_{00 \epsilon}
+2 H \epsilon_{\mu \delta \varphi} \pro_0 q_{\mu \varphi 0}
- 4 H^2 q_{00\delta}
\nn
\\
&=&
0. \label{Eq1}
\eea
The equation represents a constraint which can be
solved exactly
\bea
&&H^2 \pro_\delta q_{00\delta}
=H \pro_0\pro_\delta Q_\delta, \label{eq00d}  \\
&&q_{00 \delta}^{\rm tr}
= \dfrac{1}{\Delta+4 H^2}
\pro_0
\big(-2 \epsilon_{\delta \mu\phi} \pro_\mu Q_\phi^{\rm tr}+
4 H Q_\delta^{\rm tr}\big). \label{sol1}
\eea
The constraint allows to express
the field $q_{00\delta}$ in terms of $Q_\alpha$.
Notice that we cannot impose gauge fixing condition to eliminate
the field $Q_\alpha^{\rm l}$ since it is gauge invariant under the
Lorentz gauge transformation.
In Eq. (\ref{eq00d}) we keep $H$-terms explicitly to show that
this constraint vanishes identically in the limit $H\rightarrow 0$.
In further we will assume that $H$ is a small parameter
to justify our perturbative analysis of equations of motion.

The equation  $\delta {\cal L} / \delta q_{0\gamma\delta}$ contains
a part with time derivatives of first order.
It is convenient to use the Lorentz gauge freedom and
impose a gauge fixing condition which makes these terms vanished
\bea
&&
(\alpha+\gamma) \pro_\mu q_{\mu\gamma\delta}
- \gamma \pro_\mu\big(q_{\gamma \mu\delta}-q_{\delta \mu\gamma}\big)
\nn
\\
&&
-\alpha H\big(\epsilon_{\gamma \mu \varphi} q_{\mu\varphi \delta}
-\epsilon_{\delta \mu \varphi} q_{\mu\varphi \gamma}\big)
=0.       \label{G2}
\eea
The gauge fixing condition can be written in terms of component fields as follows
\bea
&&
2 (\alpha+\gamma) \pro_\alpha S_\alpha+\gamma\! \TS
-\dfrac{2 \alpha}{\Delta} H\pro_\alpha A_\alpha=0, \label{G2a} \\
&&
\Delta S_\alpha^{\rm tr}
-\epsilon_{\alpha \gamma\delta} \pro_\gamma A_\delta^{\rm tr}
-2 H A_\alpha^{\rm tr}=0.  \label{G2b}
\eea
The last equation allows to express the pseudo-vector field
$S_\alpha^{\rm tr}$
in terms of the physical vector field $A_\alpha^{\rm tr}$.
Since one has six gauge degrees of freedom due to the Lorentz gauge symmetry
one can impose another three gauge fixing conditions.
We will impose them later, for the present moment it is difficult
to determine which conditions should be imposed
in a consistent manner with all equations of motion.
With this, the equation $\delta {\cal L} / \delta q_{0\gamma\delta}$
results in a constraint
\bea
\dfrac{\delta {\cal L}}{\delta q_{0\gamma\delta}}
&\equiv&
(\alpha+\gamma) \Delta q_{0\gamma\delta}
+\gamma \pro_\mu
\big(\pro_\gamma q_{0\delta\mu}-\pro_\delta q_{0\gamma\mu}\big)
\nn
\\
&&
+\alpha \pro_\mu\big(\pro_\gamma q_{\delta 0\mu}
-\pro_\delta q_{\gamma 0 \mu}\big) \nn \\
&&
-2 \gamma \epsilon_{\gamma\delta\rho} \Delta Q_\rho
+2 \gamma \epsilon_{\gamma\mu\rho} \pro_\delta \pro_\mu Q_\rho
-2 \gamma \epsilon_{\delta\mu\rho} \pro_\gamma \pro_\mu Q_\rho
\nn
\\
&&
+ H\Big\{2 \alpha \pro_\gamma Q_\delta
+ 2 (\alpha+2\gamma) \epsilon_{\gamma\mu\epsilon}
\pro_\mu q_{0\delta\epsilon}
\nn
\\
&&
\hspace*{8mm}
+\gamma \epsilon_{\gamma\delta\epsilon}
\pro_\mu(q_{0\epsilon\mu}-q_{\mu\epsilon 0})
+\gamma \epsilon_{\gamma\mu\epsilon}
 \pro_\delta q_{0\epsilon \mu}
\nn
\\
&&
\hspace*{8mm}
+2(-\alpha+\gamma) \epsilon_{\gamma \mu \epsilon}
\pro_\mu q_{\delta \epsilon 0}
 -(\gamma \leftrightarrow \delta)\Big\}
\nn
\\
&&
-H^2 \Big\{2 \alpha q_{0\gamma\delta}+\alpha (q_{\gamma 0\delta}-
q_{\delta 0\gamma}) \Big\}
\nn
\\
&=&
0.    \label{Eq2}
\eea
For our purpose to determine the dynamic content
of the theory we will need the solution to this equation
up to order $H^2$,
\bea
 q_{0\gamma\delta}
&=&
-\Big(1+\dfrac{2 H^2}{\Delta}\Big)
\pro_\gamma R_\delta^{\rm tr}-\dfrac{4 \alpha-2 \gamma}{\alpha+\gamma}
H\epsilon_{\gamma\delta\alpha} R_\alpha^{\rm l}
\nn
\\
&&
-\dfrac{\alpha-\gamma}{\alpha+
\gamma}H\epsilon_{\gamma\delta\alpha} \dfrac{\pro_\alpha}{\Delta}\!\TR
+ \dfrac{1}{2} \epsilon_{\gamma\delta\alpha} Q_\alpha^{\rm tr}
+\dfrac{2 H}{\Delta} \pro_\gamma Q_\delta^{\rm tr}
\nn
\\
&&
+\dfrac{\gamma}{\alpha+\gamma}\epsilon_{\gamma\delta\alpha} Q_\alpha
+\dfrac{\alpha(\alpha+3\gamma)}{(\alpha+\gamma)^2}\dfrac{H^2}{\Delta}
\epsilon_{\gamma\delta\alpha} Q_\alpha^{\rm l}
\nn
\\
&&
-(\gamma \leftrightarrow \delta)+{\cal O}(H^{n\geq 3}).  \label{sol2}
\eea

The next equation of motion, $\delta {\cal L}/\delta q_{\nu 0\nu}$, represents
a constraint which allows to express the component field $\TR$
in terms of other fields
\bea
\dfrac{\delta {\cal L}}{\delta q_{\nu 0 \nu}}
&\equiv&
\alpha \Delta\!\TR-2 \alpha \pro_0 \pro_\mu A_\mu
\nn
\\
&&
+ H\Big\{2(3 \gamma-\alpha) \pro_0 T
\nn
\\
&&
\hspace*{8mm}
+\Big( 6 (\alpha+2\gamma)
+\dfrac{4 \gamma(2\alpha-3\gamma)}{\alpha+\gamma}\Big)
\pro_\mu Q_\mu \Big\}
\nn \\
&&
-4 H^2
\Big\{\dfrac{2\alpha-3\gamma}{\alpha+\gamma}
\Big (2 (2 \alpha-\gamma) \pro_\mu R_\mu-
(\alpha-\gamma)\!\TR \Big)
\nn
\\
&&
\hspace*{11mm}
+2 \gamma\!\TR+4\gamma \pro_\mu R_\mu \Big\}
+{\cal O}(H^{n\geq 3})
\nn
\\
&=&
0, \label{Eq3}
\eea
where we introduce a useful notation $T$ for the irreducible totally
antisymmetric part of $q_{\alpha\beta\gamma}$
\be
\begin{array}{rcl}
 q_{(\alpha\beta\gamma)}
&\equiv&
q_{\alpha\beta\gamma}+q_{\beta\gamma\alpha}+q_{\gamma\alpha\beta}
=\epsilon_{\alpha\beta\gamma} T,  \\
T
&\equiv&
\dfrac{1}{2} \epsilon_{\mu\delta\varphi} q_{\mu\delta\varphi}
=\TS+2\pro_\gamma S_\gamma.
\end{array} \label{scalT}
\ee

Let us consider the following equation of motion
\bea
\dfrac{\delta {\cal L}}{\delta q_{\nu\nu\delta}}
&\equiv&
\alpha\Big(\Box q_{\nu\nu\delta}-\pro_\delta\pro_\mu q_{\nu\nu\mu}
-\pro_\mu \pro_\nu q_{\mu\nu\delta}
\nn
\\
&&
\hspace*{5mm}
+\pro_0 \big(\pro_\nu q_{0\nu\delta}
+\pro_\delta q_{\nu\nu 0} +\pro_\nu q_{\delta 0\nu}\big)
-2 H^2 q_{\nu\nu\delta}\Big)
\nn \\
&&
+ H \Big\{
\alpha \epsilon_{\delta \epsilon \beta} \pro_\mu q_{\mu\epsilon\beta}
-3\alpha \epsilon_{\delta\beta\epsilon} \pro_\mu q_{\beta\epsilon\mu}
\nn
\\
&&
\hspace*{8mm}
-4\alpha \epsilon_{\delta \mu\epsilon} \pro_\mu q_{\nu\nu\epsilon}+
3(\alpha+2\gamma) \epsilon_{\mu\beta \epsilon}
\pro_\mu q_{\beta\epsilon\delta}
 \nn \\
&&
\hspace*{8mm}
-(2\alpha-3\gamma) \epsilon_{\beta\mu\epsilon}
\pro_\mu q_{\delta\epsilon \beta}
+\alpha \epsilon_{\delta\nu\epsilon} \pro_0 q_{\nu\epsilon 0}
\nn
\\
&&
\hspace*{8mm}
+(\alpha-3 \gamma) \epsilon_{\epsilon\mu\beta}
\pro_\delta q_{\epsilon\mu\beta}
-\alpha \epsilon_{\delta\epsilon\beta}\pro_0 q_{0\epsilon\beta}
\Big\}
\nn
\\
&=&
0,
\label{Eq4}
\eea
where $\Box \equiv -\pro_0 \pro_0 +\pro_\alpha \pro_\alpha$.
The transverse part of the equation leads to propagation equation
for the transverse
part of the vector field $A_\delta=-q_{\nu\nu\delta}/2$,
\be
\Box q_{\nu\nu\delta}-\dfrac{\pro_\delta\pro_\mu}{\Delta}
q_{\nu\nu\mu} + {\cal O}(H^{n\geq 1})=0.
\label{dynspin1}
\ee
The longitudinal part of the equation at the lowest order $H^0$
coincides with the lowest order part of the Eq.(\ref{Eq3}), so that
a nontrivial part of the equation appears at the next order in $H$:
\bea
&&
H\Big \{(\alpha+\gamma) \Delta T -(\alpha-3 \gamma)
\pro_0\pro_0 T+
\dfrac{\alpha}{\gamma}(\alpha+5\gamma)\pro_0 W\Big \}
\nn
\\
&&
+{\cal O}(H^{n\geq 2})=0,    \label{TW}
\eea
where the field $W$ corresponds to a scalar irreducible
part of $q_{0\gamma\delta}$,
$$
W\equiv \dfrac{1}{2} \epsilon_{\alpha\beta\gamma}
\pro_\alpha q_{0\beta\gamma}
=\dfrac{2\gamma}{\alpha+\gamma} \pro_\alpha Q_\alpha +{\cal O}(H).
$$
As we will see below, the fields $T$ and $W$
represent propagating scalar modes corresponding to
the longitudinal field components $S_\alpha^{\rm l}, Q_\alpha^{\rm l}$.
Notice, that one has arbitrariness in choosing a set of independent
field variables in the theory. The equation (\ref{TW})
contains fields $\TS, S_\alpha^{\rm l}$ which satisfy the gauge fixing
condition (\ref{G2a}) including the field $A_\alpha^{\rm l}$.
So that, it is appropriate (and consistent with all other equations of motion)
to treat the constraint (\ref{TW}) as a non-linear
equation for $A_\alpha^{\rm l}$.

Let us now consider
the equation $\delta {\cal L}/\delta q_{\beta 0\delta}$
\bea
\dfrac{\delta {\cal L}}{\delta q_{\beta 0 \delta}}
&\equiv&
\alpha \Delta q_{\beta 0\delta}
+\gamma \Delta \big(q_{\beta 0\delta}-q_{\delta 0 \beta}\big)
-\alpha \pro_\mu\pro_\beta q_{\mu 0\delta}
\nn
\\
&&
-\alpha \pro_0\pro_0 (q_{\beta 0\delta}-q_{\delta 0 \beta})
-\gamma\Delta q_{0\beta\delta} +\alpha \pro_0\pro_\beta q_{00\delta}
\nn
\\
&&
-\alpha \pro_0\pro_\delta q_{00\beta}
-\alpha \pro_0\pro_\mu q_{\delta \mu\beta}
+\gamma \pro_0\pro_\mu q_{(\mu\beta\delta)}
\nn
\\
&&
+\gamma \pro_\mu
\Big(-\pro_\beta q_{\mu 0\delta}
+\pro_\delta q_{\mu 0\beta}+\pro_\beta q_{\delta 0\mu}
-\pro_\delta q_{\beta 0\mu}
\Big)
\nn \\
&&
+\alpha \pro_\mu\pro_\delta q_{0\mu\beta}
-\gamma \pro_\mu\pro_\delta q_{0\mu\beta}
+\gamma\pro_\mu \pro_\beta q_{0\mu\delta} \nn \\
&&
+H\Big\{
2 \alpha \epsilon_{\beta\delta\epsilon} \pro_0 q_{00\epsilon}
-\alpha \epsilon_{\delta \epsilon\mu}\pro_0q_{\epsilon \mu\beta}
\nn
\\
&&
\hspace*{8mm}
+\gamma \epsilon_{\delta\epsilon\mu} \pro_0 q_{(\epsilon\mu\beta)}
+2(\alpha+2\gamma) \epsilon_{\delta\mu\epsilon}
\pro_\mu q_{\beta\epsilon\mu}
\nn
\\
&&
\hspace*{8mm}
+2\gamma \epsilon_{\delta\mu\epsilon} \pro_\mu q_{\epsilon 0\beta}
-(\alpha+2\gamma) \epsilon_{\delta \mu \phi}\pro_\beta q_{\mu\phi 0}
\nn
\\
&&
\hspace*{8mm}
+(\alpha+2\gamma)\epsilon_{\beta \delta\epsilon}
\pro_\mu q_{\mu\epsilon 0}
+(\alpha-2 \gamma)\epsilon_{\beta\delta\epsilon}
\pro_\mu q_{0\epsilon\mu}
\nn
\\
&&
\hspace*{8mm}
+2\gamma\epsilon_{\beta\mu\epsilon} \pro_\mu q_{\delta 0\epsilon}
+2(\alpha-2 \gamma)\epsilon_{\delta\mu\epsilon}
\pro_\mu q_{0\epsilon\beta}
\nn
\\
&&
\hspace*{8mm}
+2\gamma \epsilon_{\beta\mu\epsilon}\pro_\mu q_{0\epsilon\delta}
-\gamma\epsilon_{\delta\phi\mu} \pro_\beta q_{0\phi\mu}\Big \}
\nn \\
&&
+H^2
\Big\{
-(3\alpha+2\gamma)q_{\beta 0\delta}+
(\alpha-2\gamma)\delta_{\beta\delta} q_{\nu 0\nu}
\nn
\\
&&
\hspace*{10mm}
-\alpha
q_{0\beta\delta}
\Big\}
\nn
\\
&=&
0. \label{Eq5}
\eea
To solve this equation for all its irreducible parts one has to
take into account terms up to order $H^2$ because
some irreducible components of this equation vanish at the lower order
expansion in $H$.
Let us start with the equation obtained
by contraction with the antisymmetric tensor
$\epsilon_{\alpha\beta\delta}$.
This equation produces two constraints.
The first one corresponds to the longitudinal projection of
the contracted equation,
and it can be simplified to the following constraint
\bea
&&
-2 \gamma \pro_\alpha W -\alpha \pro_0 \pro_\alpha\!\TS
+2\gamma \pro_0 \pro_\alpha T
\nn
\\
&&
+2 (\alpha+2 \gamma) \Delta Q_\alpha^{\rm l}
-4 \alpha \pro_0\pro_0 Q_\alpha^{\rm l}
\nn
\\
&&
+ {\cal O}(H^{n\geq 1})=0. \label{S5.1}
\eea
This equation provides propagation equation for $Q_\alpha^{\rm l}$.
The transverse part of the antisymmetrized equation (\ref{Eq5})
vanishes at the lowest order $H^0$ and leads to a
non-linear relationship between fields $R_\delta^{\rm tr}$
and $A_\delta^{\rm tr}$
\bea
&&
H\pro_0 A_\alpha^{\rm tr}
+ H^2 \epsilon_{\alpha\beta\delta} \pro_\beta R_\delta^{\rm tr}-
\Big(2+\dfrac{\gamma}{\alpha}\Big)H^2 Q_\alpha^{\rm tr}
\nn
\\
&&
+{\cal O}(H^{n\geq 3})=0. \label{RA}
\eea

Let us now consider the equation (\ref{Eq5}) symmetrized over its indices.
The divergence of the symmetrized part
$\pro_\beta \delta {\cal L}/\delta q_{\{\beta 0 \delta\}}$
implies two equations.
First one does not vanish only at order $H^1$, and it produces the same
constraint as (\ref{S5.1}). The second equation is
\bea
&&
(2\alpha+3 \gamma) H \Delta Q_\delta^{\rm tr}
\nn
\\
&&
-2 (3\alpha+\gamma)H^2
\Big(\epsilon_{\delta \gamma\rho}\pro_\gamma Q_\rho^{\rm tr}
+\Delta R_\delta^{\rm tr}\Big)
\nn
\\
&&
+{\cal O}(H^{n\geq 3})=0.
\eea
Due to relationship (\ref{RA}) between the fields $R_\alpha^{\rm tr}$
and $ A_\alpha^{\rm tr}$
the last constraint implies, in general, a vanishing condition for both fields
$R_\alpha^{\rm tr}$, $A_\alpha^{\rm tr}$ and absence of any propagating
modes in the model.
There is only one special case where our model admits dynamical vector field,
namely, we choose a condition on the parameters
\be
\gamma=-3 \alpha  \label{gamma3}
\ee
which excludes the field  $R_\alpha^{\rm tr}$ from the equation.
With this the field $A_\alpha^{\rm tr}$ remains dynamical.
Our careful analysis shows that this condition is consistent with
all other equations of motion and with Noether identities.
Notice, the constraint on the parameters is exactly the same as in the case of
the model of the gravity with contortion in the presence of constant curvature
space-time background \cite{IJMPA2010}. This is an unexpected result
because we have different equations of motion in the models
with flat and non-flat metric.

At this moment we can choose remaining three gauge fixing conditions in
a suitable manner.
From the last constraint and previous solutions to the equations of motion
one can verify that the fields $Q_\delta^{\rm tr}$ and $R_\delta^{\rm l}$
do not affect the solution structure in principle. It is convenient
to choose vanishing conditions for $Q_\delta^{\rm tr}$ and
$R_\delta^{\rm l}$
which are consistent with equations of motion and simplify further calculations.
So that, from now on we impose the gauge fixing conditions
\be
Q_\delta^{\rm tr}= 0, \qquad
R_\delta^{\rm l}= 0.
\ee
With the previously imposed gauge conditions (\ref{G2})
the Lorentz gauge symmetry has been fixed completely.

The remaining equation  corresponding to the traceless and transverse
part of the
equation $\delta {\cal L}/\delta q_{\{\beta 0 \delta\}}$
gives a relationship for spin two modes
\bea
\Delta \TTR_{\beta\delta}
&=&\dfrac{1}{2} \pro_0
\Big(\epsilon_{\alpha\delta\rho} \pro_\alpha\!\!\TTS_{\beta\rho}
+\epsilon_{\alpha\beta\rho} \pro_\alpha\!\!\TTS_{\delta\rho}\!\Big)
\nn
\\
&&
+ {\cal O}(H).  \label{spin2rel}
\eea

The last equation of motion
is given by $\dfrac{\delta {\cal L}}{\delta q_{\beta\gamma\delta}}$.
It is convenient to rewrite this equation in a dual form
\bea
\Phi_{\alpha\beta}
&\equiv&
\epsilon_{\alpha\gamma\delta} \dfrac{\delta {\cal L}}
{\delta q_{\beta\gamma\delta}} \nn \\
&=&
\alpha \Box \starq_{\beta\alpha}
+ \gamma \delta_{\alpha\beta} \Box \starq_{\nu\nu}
+(\gamma-\alpha) \pro_\beta\pro_\mu \starq_{\mu\alpha}
\nn
\\
&&
-\gamma\Big(\Delta \starq_{\beta\alpha}
-\pro_\alpha\pro_\rho \starq_{\beta\rho}
+\delta_{\alpha\beta} \pro_\mu\pro_\nu \starq_{\mu\nu}
+\pro_\alpha\pro_\beta \starq_{\nu\nu}\Big)
\nn
\\
&&
 -(\alpha-\gamma)\epsilon_{\alpha\gamma\rho} \pro_\gamma
\pro_\mu q_{\rho\mu\beta}
+\alpha \epsilon_{\alpha\gamma\rho}
\pro_0\pro_\gamma q_{\rho 0 \beta}
\nn
\\
&&
-2 \gamma\pro_0
\big(\delta_{\alpha\beta}\pro_iQ_i-\pro_\beta Q_\alpha\big)
-2\gamma\pro_0\pro_\beta Q_\alpha
\nn
\\
&&
+\dfrac{1}{2}(\alpha+\gamma)
\epsilon_{\alpha\gamma\delta}\pro_0\pro_\beta q_{0\gamma\delta}
+\gamma\epsilon_{\alpha\gamma\rho}\pro_0\pro_\gamma q_{0\rho\beta}
\nn \\
&&
+ H
\Big\{
2 \gamma \pro_\mu q_{\mu\alpha\beta}
-(\alpha+2\gamma) \pro_\mu q_{\beta\alpha\mu}
-\alpha \pro_\beta q_{\nu\nu\alpha}
\nn
\\
&&
\hspace*{8mm}
+2(\alpha-\gamma)\pro_\alpha q_{\nu\nu\beta}
-\alpha \pro_\mu q_{\alpha\mu\beta}-\alpha \pro_0 q_{0\alpha\beta}
\nn \\
&&
\hspace*{8mm}
+(\alpha-2\gamma) \delta_{\alpha\beta} \pro_0 q_{\nu\nu 0}
-\alpha \pro_0 q_{\beta\alpha 0}
\nn
\\
&&
\hspace*{8mm}
-(\alpha-2\gamma)\delta_{\alpha\beta}\pro_\mu q_{\nu\nu\mu}
\nn
\\
&&
\hspace*{8mm}
+2\gamma\pro_\mu \big(\epsilon_{\mu\beta\alpha} \starq_{\nu\nu}
-\epsilon_{\mu\beta\gamma} \starq_{\gamma\alpha}\big)\Big\} \nn \\
&&
+H^2 \Big\{
4\gamma(\starq_{\beta\alpha}-\starq_{\alpha\beta})
-(4\gamma+\alpha)\epsilon_{\alpha\beta\gamma} q_{\nu\nu\gamma}
\Big\}, \label{Eq6}
\eea
where $\starq_{\beta\alpha}\equiv
\dfrac{1}{2} \epsilon_{\alpha\gamma\delta} q_{\beta\gamma\delta}$.
The trace part of the equation,
$\Phi_{\alpha\alpha}$, yields an equation which can be
simplified using the condition $\gamma=-3 \alpha$
\be
-4 \Box \pro_\alpha S_\alpha
+2 \Delta \pro_\alpha S_\alpha
-3 \pro_0 \pro_\alpha Q_\alpha +{\cal O}(H)=0. \label{eq38}
\ee
The eqs.(\ref{eq38}) and (\ref{S5.1}) imply that the longitudinal components
of the vector fields $S_\alpha$, $Q_\alpha$ become propagating. Defining
a scalar field corresponding to the longitudinal component
of $S_\alpha$
\bea
\psi&=&-\dfrac{2}{3}\pro_\alpha S_\alpha,
\eea
one can rewrite the equations of motion as follows
\be
\begin{array}{rcl}
&&
\Box \pro_\alpha Q_\alpha
+\Delta \big(\pro_\alpha Q_\alpha+\pro_0 \psi\big)
+{\cal O}(H) = 0, \\
&&
\Box \psi-\pro_0 \big(\pro_\alpha Q_\alpha+\pro_0 \psi\big)
+{\cal O}(H)= 0.
\end{array}
\label{eq46}
\ee
Explicit expressions for propagating solutions to these equations will be
given in the next section.

The remaining equations of motion corresponding to the vector irreducible
parts of $\Phi_{\alpha\beta}$ do not produce new independent equations.
The irreducible part of the equation
$\epsilon_{\delta\alpha\beta}\Phi_{\alpha\beta}$ coincides
with (\ref{Eq4}).
The divergence of the equation (\ref{Eq6}), $\pro_\alpha\Phi_{\alpha\beta}$,
reproduces the same propagating equation
for $A_\mu^{\rm tr}$ as in (\ref{dynspin1}) and the constraint (\ref{S5.1}).
The divergence of the equation (\ref{Eq6}) with respect to the second
index, $\pro_\beta\Phi_{\alpha\beta}$, reflects the Noether identity
structure. One can verify that the transverse part of this equation leads to
a nontrivial equation at order $H^{n\geq 1}$
\be
\alpha H \Big\{ \Delta A_\alpha^{\rm tr}-H \epsilon_{\alpha\beta\gamma}
\pro_0 \pro_\beta R_\delta^{\rm tr} \Big\}
+{\cal O}(H^{n\geq 3})= 0,
\ee
which is consistent with the constraint (\ref{RA}).
The longitudinal part of the equation $\pro_\beta\Phi_{\alpha\beta}$
can be simplified by using the constraint (\ref{Eq3}),
\be
H \dfrac{\pro_0 \pro_0}{\Delta} \pro_\alpha A_\alpha
+ {\cal O}(H^{\geq 2})=0. \label{long1}
\ee
The component field $A_\alpha^{\rm l}$ has been already defined
by the Eq. (\ref{TW}).
The equation (\ref{long1}) does not represent a new independent equation
but reflects the structure of the solution of (\ref{TW}).
Namely, the equation contains second order time derivative which indicates
on possibility of existence of wave like (soliton) solutions
for $A_\alpha^{\rm l}$ in the full non-linear theory
beyond the linearized approximation given by decomposition (\ref{split}).

The last irreducible component of the equation $\Phi_{\alpha\beta}$
is given by its symmetric traceless part.
Substituting the irreducible field $\TTR_{\alpha\beta}$
from (\ref{spin2rel}) and using a useful identity
\be
\TTS_{\alpha\beta}+\epsilon_{\alpha\gamma\delta} \epsilon_{\beta\nu\rho}
 \dfrac{\pro_\gamma \pro_\nu}{\Delta} \TTS_{\delta\rho}= 0,
\ee
one results in the following equation at order $H^2$,
\be
H^2 \Big\{(\alpha+\gamma)
\dfrac{\pro_0\pro_0}{\Delta}\!\TTS_{\alpha\beta}\!\Big\}
+{\cal O}(H^{n \geq 3})=0. \label{spin2}
\ee
The equation contains second order time derivative,
that means there might be spin two propagating
solution like soliton due to non-linearity of the initial
equations of motion.
The difference of the equations of motion for $\TTS_{\alpha\beta}$
in the case of constant torsion background
and in the case of the gravitational
space-time background \cite{IJMPA2010} is that Eqn.
(\ref{spin2}) does not represent a standard D'Alembert
equation due to the absence of a term proportional
to $H^2\!\TTS_{\alpha\beta}$ which would produce the D'Alembert equation.

Finally, we have demonstrated that the Lorentz gauge theory
with Lagrangian (\ref{startL})
with parameters $\gamma=-3 \alpha, \beta=0$ admits
two transverse propagating modes for the vector field
$A_\alpha^{\rm tr}$ and two scalar propagating modes $Q_\alpha^{\rm l}$,
$S_\alpha^{\rm l}$.
The spin one mode $A_\alpha^{\rm l}$ and spin two mode $\TTS_{\alpha\beta}$
might have propagating modes
only due to non-linear structure of full equations of motion.
Our result that the Lagrangian has exactly the same structure,
$\gamma=-3\alpha,\beta=0$, as the Lagrangian for the gravity
with torsion in the presence of the background metric \cite{IJMPA2010}
confirms that the propagating spin one mode exists independently on
the background metric at hand and it is a feature of
the Lorentz gauge model itself.

\section{Positive definiteness of the Hamiltonian}

Lorentz gauge theories with quadratic $R^2$ type Lagrangian
suffer from the non-positiveness problem of the Hamiltonian
which has origin in the non-compact structure of the Lorentz group.
This leads to the problem of defining a stable vacuum
in quantum theory.
Let us consider this problem starting with the
free Lagrangian (\ref{startL}). Using solutions from the
previous section one can express all components of contortion $q_{acd}$
in terms of three independent
fields $A_\alpha^{\rm tr}$, $S_\alpha^{\rm l}$ and $Q_\alpha^{\rm l}$,
\be
{\cal L}^{(2)}
=
\dfrac{1}{2}
\Big[
A^{\rm tr}_\alpha \Box A^{\rm tr}_\alpha
+(\pro_\alpha Q_\alpha^{\rm l}+\pro_0 \psi)^2
+\psi \Box \psi-Q_\alpha^{\rm l} \Box Q_\alpha^{\rm l}
\Big ]. \label{L2}
\ee
Since the vector fields $A_\alpha^{\rm tr}$ and $S_\alpha^{\rm tr}$
are related by the Eq. (\ref{G2b}), one can treat the scalar field
$\psi=-\frac{2}{3}\pro_\alpha S_\alpha$ as a longitudinal
component of $S_\alpha$,
 or as a dual longitudinal component of the field $A_\alpha^{\rm tr}$.
The field $Q_\alpha^{\rm l}$ originates from
the contortion part $q_{\alpha 0 \delta}$
which corresponds to boost generators of the Lorentz group.
The terms with $Q_\alpha^{\rm l}$ in the Lagrangian
are potentially dangerous since they may give negative
energy contribution destabilizing the vacuum.
We concentrate on a part of the total Hamiltonian which includes
the scalar modes $\psi$ and $Q_\alpha^{\rm l}$.
The Hamiltonian is defined in a standard manner
\be
{\cal H}(Q_\alpha^{\rm l},\psi)
=\dfrac{1}{4} (\pi-\pro_\alpha Q_\alpha^{\rm l})^2-
\dfrac{1}{2}\pi_\alpha^2 +\dfrac{1}{2}(\pro_\alpha \psi)^2
-(\pro_\alpha Q_\alpha^{\rm l})^2, \label{hamilt}
\ee
where canonical momentums $\pi$ and $\pi_\alpha$ are defined by
\renewcommand{\arraystretch}{2.2}
\be
\begin{array}{rcl}
\pi
&=&
\dfrac{\pro {\cal L}}{\pro \pro_0 \psi}
=2 \pro_0 \psi +\pro_\alpha Q_\alpha^{\rm l}, \\
\pi_\alpha
&=&
\dfrac{\pro {\cal L}}{\pro \pro_0 Q_\alpha^{l}}=-\pro_0  Q^{\rm l}_\alpha.
\end{array}
\ee
\renewcommand{\arraystretch}{1.6}

\vspace*{-4mm}
\noindent
Notice that the fields $\psi$ and $Q_\alpha^{\rm l}$ have correct
canonical dimension and they are treated as initial independent field variables.
We will solve the Euler-Lagrange equations of motion for the fields
$\psi$, $Q_\alpha^{\rm l}$, (\ref{eq46}), in lowest order approximation.
For a convenience let us rewrite the equations (\ref{eq46})
in the following form
\be
\begin{array}{rcl}
&&2 \pro_0^2 \psi-\Delta \psi +\pro_0 \pro_\alpha Q_\alpha^{\rm l}=0, \\
&&  \pro_0^2 Q_\alpha^{\rm l}- 2 \Delta Q_\alpha^{\rm l}
-\pro_0 \pro_\alpha \psi=0.
\end{array}
\label{syseq}
\ee
The system of equations (\ref{syseq}) cannot be factorized into decoupled
equations. Let us consider possible solutions in the form of plane waves
\be
\begin{array}{rcl}
\psi(k)
&=&
b(k) {\rm e}^{i(-\vec k \vec  x+ k_0 t)}, \\
Q_\alpha^{\rm l}(k)
&=&
c_\alpha (k) {\rm e}^{i(-\vec k \vec  x+ k_0 t)},
\end{array}
\ee
where $\vec k \vec x =k_\alpha x_\alpha$.
Substitution of the plane waves into (\ref{syseq})
gives a system of homogeneous equations
which has a nontrivial solution
if the following
characteristic equation is satisfied
\be
(k_0^2-\vec k^2)^2=0. \label{dispeq}
\ee
The equation is degenerated and it implies the dispersion relationship
\be
k_0 = \pm \omega, \qquad \mbox{with}
\qquad \omega \equiv  \sqrt {\vec k^2}. \label{disprel}
\ee
The coefficient functions $b, c_\alpha$ are related by the following
equation
\be
c_\alpha (k) = \dfrac{k_0 k_\alpha}{\omega^2} b(k).
\ee
The corresponding solution for $\psi, Q^{\rm l}_\alpha$ can be
written as a sum of positive and negative frequency
modes
\renewcommand{\arraystretch}{2.4}
\be
\begin{array}{rcl}
\psi(\vec{x},t)
&=&
\displaystyle
\!\int\!\dfrac{{\rm d}^3 \vec k}{(2 \pi)^4} b^+(\vec k)
{\rm e}^{i(-\vec k \vec x+\omega t)}
\\
&&
\hspace*{-4mm}
\displaystyle
+\!\int\!\dfrac{{\rm d}^3 \vec k}{(2 \pi)^4} b^- (\vec k)
{\rm e}^{-i(\vec k \vec x+\omega t)},
\\
\vec Q^{\rm l}(\vec{x},t)
&=&
\displaystyle
\!\int\!\dfrac{{\rm d}^3 \vec k}{(2 \pi)^4}
\dfrac{b^+(\vec k) \vec k}{\omega}
{\rm e}^{i(-\vec k \vec x+\omega t)}
\\
&&
\hspace*{-4mm}
\displaystyle
-\!\int\!\dfrac{{\rm d}^3 \vec k}{(2 \pi)^4}
\dfrac{b^- (\vec k) \vec k}{\omega}
{\rm e}^{-i(\vec k \vec x+\omega t)}.
\end{array}
\ee
Using the solutions and calculating the canonical momentums $\pi$ and
$\pi_\alpha$, one can easily check the identities
\renewcommand{\arraystretch}{2.0}
\be
\begin{array}{rcl}
&&\dfrac{1}{4}
(\pi-\pro_\alpha Q_\alpha^{\rm l})^2
-(\pro_\alpha Q_\alpha^{\rm l})^2 =0, \\
&&-\dfrac{1}{2}\pi_\alpha^2 +\dfrac{1}{2}(\pro_\alpha \psi)^2 =0,
\end{array}
\ee
\renewcommand{\arraystretch}{1.6}
which imply immediately that the Hamiltonian (\ref{hamilt})
vanishes identically.

Since the equation (\ref{dispeq}) is degenerated
the general solution to the equations of motion (\ref{syseq}) includes another
couple of wave like solutions. Fourier modes of the solutions
can be found in the form which is suitable in further making
Lorentz invariant decomposition into positive and negative frequency parts
\be
\begin{array}{rcl}
\psi(k)
&=&
 (\vec k \vec x +k_0 t) a(k) {\rm e}^{i(-\vec k \vec  x+ k_0 t)}, \\
\vec Q^{\rm l}(k)
&=&
\Big((\vec  k \vec x +k_0 t) \vec a(k)+ i \vec d(k)\Big)
{\rm e}^{i(-\vec k \vec  x+k_0 t)}.
\end{array}
\ee
Substituting this ansatz into equations of motion
produces the same dispersion relation (\ref{dispeq})
and following relations for the coefficient functions
\be
\begin{array}{rcl}
\vec a
&=&
\dfrac{ k_0 \vec k}{\omega^2} a, \\
\vec d
&=&
- 6 \vec a= -\dfrac{6 k_0 \vec k}{\omega^2} a.
\end{array}
\ee
The general solution for $\psi$ and $\vec Q^{\rm l}$ can be represented
as Fourier integral over all momentum $\vec k, k_0$.
Performing integration over $k_0$ using the dispersion relation (\ref{disprel})
leads to the final expressions
\renewcommand{\arraystretch}{2.4}
\be
\begin{array}{rcl}
\psi(\vec{x},t)
&=&
\displaystyle
\!\int\!\dfrac{{\rm d}^3 \vec k}{(2 \pi)^4}
(\vec k \vec x+\omega t) a^+(\vec k) {\rm e}^{i(-\vec k \vec x+\omega t)}
\\
&&
\displaystyle
\hspace*{-4mm}
+\!\int\!\dfrac{{\rm d}^3 \vec k}{(2 \pi)^4}
(\vec k \vec x-\omega t)a^- (\vec k) {\rm e}^{-i(\vec k \vec x+\omega t)}, \\
\vec Q^{\rm l}(\vec{x}, t)
&=&
\displaystyle
\!\int\!\dfrac{{\rm d}^3 \vec k}{(2 \pi)^4}
\Big((\vec k \vec x+\omega t)-6i\Big)
\dfrac{ \vec k}{\omega} a^+(\vec k) {\rm e}^{i(-\vec k \vec x+\omega t)}
\\
&&
\displaystyle
\hspace*{-4mm}
-\!\int\!\dfrac{{\rm d}^3 \vec k}{(2 \pi)^4}
\Big((\vec k \vec x-\omega t)-6i\Big)
\dfrac{ \vec k}{\omega}a^- (\vec k)  {\rm e}^{-i(\vec k \vec x+\omega t)}.
\label{finsol}
\end{array}
\ee
\renewcommand{\arraystretch}{1.6}

\vspace*{-5mm}
\noindent
As usual, the Fourier functions $a^\pm(\vec k), b^\pm(\vec k)$ turn
into creation and annihilation operators during quantization procedure.
It is convenient to split the Hamiltonian ${\cal H}(Q_\alpha^{\rm l},\psi)$
into two parts
\bea
&&
{\cal H}= {\cal H}_1+ {\cal H}_2, \\
&&
{\cal H}_1 \equiv  \dfrac{1}{4}
(\pi-\pro_\alpha Q_\alpha^{\rm l})^2
-(\pro_\alpha Q_\alpha^{\rm l})^2, \nn \\
&&
{\cal H}_2 \equiv -\dfrac{1}{2}\pi_\alpha^2 +\dfrac{1}{2}(\pro_\alpha
\psi)^2.
\nn
\eea
This allows to separate contributions $P_{01},~P_{02}$
of the fields $Q_\alpha^{\rm l},~\psi$ to the total
energy functional
\be
P_0=\!\int\! {\rm d}^3 x {\cal H}=P_{01}+P_{02}.
\ee
Substituting the solution (\ref{finsol}) into the last equation
and performing integration over configuration space $\vec x$ and
one of two momentum $\vec k, \vec k'$ corresponding to Fourier components
of $\psi, \vec Q^{\rm l}$ one can
verify that the contributions from
the fields $Q_\alpha^{\rm l}$ and $\psi$
are mutually canceled due to following relations
\bea
P_{01}^{+-}
&=&
\!\int\!\dfrac{{\rm d}^3 \vec k}{(2 \pi)^4}
48 \omega^2a^+(\vec k) a^-(-\vec k),
\nn
\\
P_{02}^{+-}&=&-P_{01}^{+-},
\nn \\
P_{01}^{++}
&=&
-\!\int\!\dfrac{{\rm d}^3 \vec k}{(2 \pi)^4}
\Big(8\omega^2 (3+i \omega t)\Big)a^+(\vec k) a^+(-\vec k)
{\rm e}^{2i\omega t},
\nn
\\
P_{02}^{++}&=&
-P_{01}^{++},  \\
P_{01}^{--}
&=&
-P_{02}^{--}. \nn
\eea
So that, the total contribution of the scalar modes to the energy functional
vanishes identically.

It is worth to stress that the mutual exact cancellation of all contributions
of scalar modes in the energy functional is not occasional. This indicates to
presence of an additional symmetry in the defining equations (\ref{syseq}).
It is easy to see such a symmetry in a simple case
of $1+1$ dimensional space-time. After changing variable
$\pro_x Q_x^{\rm l} \rightarrow \pro_0 \chi$ the system of
equations
(\ref{syseq}) can be rewritten in the form
\be
\begin{array}{rcl}
&&2 \pro_0^2 \psi-\pro_x^2 \psi +\pro_0 \pro_0 \chi=0, \\
&& \pro_0^2 \chi - 2 \pro_x^2 \chi -\pro_x^2 \psi=0.
\end{array}
\label{syseq11}
\ee
It is clear that the system is invariant under the following symmetry
transformations
\be
x \leftrightarrow \pm t, \qquad  \psi \leftrightarrow \pm \chi.
\ee
Due to this, energy contributions of scalar modes in (\ref{L2})
are mutually canceled. We expect that in $3+1$ dimensions there should be
a similar symmetry which provides the positive definite energy on mass shell.

\section{Discussion}

We have studied the dynamic content of the class of Lorentz gauge theories
admitting topological phase in the gravitational sector.
It has been shown that in the special choice of the parameters
$\alpha=1, \beta=0, \gamma=-3$ the corresponding model
possesses dynamical contortion. Surprisingly, the existence of
propagating modes for spin one and zero contortion component fields
is provided by the same Lagrangian in both cases,
in presence of constant gravitational background and
in presence of constant contortion background field.
Additional spin one and spin two propagating modes
may appear only due to full non-linear structure
of the equation of motion.
At the lowest order of perturbation theory we have proved
that the Hamiltonian is positively defined. This implies that
perturbative quantization can be performed straightforward.
In practical calculation it is much more convenient to
use the covariant quantization formalism based on functional integral.
The quantization can be performed straightforward in a similar
manner as in \cite{IJMPA2010}. It has been proved that quantum gravity model
with a general $R^2$ type Lagrangian is renormalizable \cite{utidewit,stelle,
fradkin-ts81,avramidi-barvinsky}.
Since the initial Lagrangian (\ref{startL})
is expressed in terms of gauge invariant tensors
and there is no dimensional coupling constants,
the proposed model of Lorentz gauge gravity belongs to
renormalizable type.

The important question is whether our model leads to a quantum vacuum
condensate of torsion which can provide generation of the Einstein term in
the effective action of gravity. This mechanism is similar
to dynamical symmetry breaking in quantum chromodynamics
where one has a gluon condensate while the gluon itself is not
observable at classical level. The possibility that
torsion may not be observable as a classical object
was pointed out in \cite{hanson}.
Generation of the vacuum torsion condensate due to appearance of a
non-trivial minimum in the quantum effective potential
would lead to an effective Einstein gravity.
Suppose the vacuum condensate has a Lorentz invariant form
$\langle{\cal R}_{abcd}\rangle
=M^2 (\eta_{ac}\eta_{bd}-\eta_{ad}\eta_{bc})$.
Substituting it into the initial classical Lagrangian (\ref{startL})
one can obtain the lowest order terms in the effective Lagrangian
of gravity
\bea
{\cal L}_{\rm eff}
&=&
-\dfrac{3}{4} M^4+\dfrac{3}{8} M^2 \hat R-\dfrac{1}{32}
\Big(\hat R_{abcd}^2
-4 \hat R_{ab}^2+\hat R^2 \Big)
\nn
\\
&&
+ {\cal O}(\hat R^{n\geq 3}), \label{effL}
\eea
where the terms quadratic in Riemann curvature represent the
integral density for the Euler characteristic
\be
\chi=\dfrac{1}{8\pi^2}\!\int\!{\rm d}^4x\sqrt{-g}
\Big(\hat R_{abcd}^2-
4 \hat R_{ab}^2 +\hat R^2\Big).
\ee
To provide the correct sign of the Einstein term
the condensate parameter $M^2$ should be negative.
This is opposite to the case of the gravity model with Yang-Mills type
Lagrangian \cite{CQG2008} where
the Einstein-Hilbert term and cosmological constant are induced
when the torsion condensate corresponds to a positive constant
Riemann-Cartan curvature, i.e.\ $M^2>0$.
Notice that the cosmological term proportional to $M^4$
is reproduced with a correct sign. Another feature of our model is that
the Euler characteristic enters the effective Lagrangian with
a negative sign. The corresponding vacuum to vacuum transition amplitude
is proportional to (in Euclidean space-time )
\be
\langle 0|0\rangle \simeq {\rm e}^{-S_E}
={\rm e}^{+\frac{\scriptstyle\pi^2}{4}\chi}.
\ee
It is reasonable to consider summation over all topologies
of the four dimensional manifolds described by fiber bundles with a compact
two dimensional base space. In that case the Euler characteristic
is determined by the genus $g$ of the base space,
$\chi = 2 - 2 g$, and the total vacuum-vacuum amplitude remains finite
after summation over all topologies.

The possibility that the Lorentz gauge gravity may have a positive
definite classical Hamiltonian bounded from below implies
that torsion can be observable not only in the form of quantum vacuum
condensate but also in the form of a classical configuration.
This implies an attractive possibility that torsion can be responsible for
the cold dark matter since it does not interact to photon in
minimal interaction scheme.
The quantum properties and possible physical implications of our model
will be considered in a separate paper.

\acknowledgments

One of the authors (DGP) thanks Y.M. Cho for suggesting the problem and
E.N. Tsoy for useful discussions. The author (DGP) acknowledges
Y. Kim and the APCTP staff for kind hospitality during his visit.
The work of D.G. Pak is supported by CAS (Contract No. 2011T1J31)
and by UzFFR (Grant No. F2-FA-F116).
Y. Kim and T. Tsukioka acknowledge the Max Planck Society (MPG),
the Korea Ministry of Education, Science and Technology (MEST),
Gyeongsangbuk-Do and Pohang City for the support of the
Independent Junior Research Group at APCTP.

\end{document}